\documentclass[prd,showpacs,preprintnumbers,amsmath,amssymb,superscriptaddress]{revtex4}
\usepackage{graphics,color}
\definecolor{rd}{rgb}{.99,0,0}
\definecolor{ma}{rgb}{.66,0,.66}
\definecolor{gn}{rgb}{0.4,.99,0.4}
\definecolor{be}{rgb}{0,0,.99}

\font\sm=cmr9

\def\Kg{{K}}
\def\calC{{\cal C}}
\def\calT{{\cal T}}
\def\calU{{\cal U}}
\def\calI{{\cal I}}
\def\calL{{\cal L}}

\def\gammag{{\gamma}}
\def\sigmag{{\sigma}}

\def\Delg{{\Delta}}

\def\calE{{\cal E}}
\def\ig{{i}}

\def\etag{{\eta}}
\def\perpg{{\perp}}
\def\epsg{{\epsilon}}

\def\smM{{\sm M}}
\def\smG{{\sm G}}

\def\omeg{{\omega}}
\def\kappar{{\kappa}}
\def\Tr{{T}}

\def\Jr{{J}}
\def\fr{{f}}

\def\Lambdar{{\Lambda}}
\def\Lr{{L}}
\def\Br{{B}}
\def\Hr{{H}}

\def\chir{{\chi}}
\def\calR{{\cal R}}
\def\nab{{\nabla}}
\def\gb{{g}}
\def\phib{{\phi}}
\def\hb{{h}}
\def\xib{{\xi}}

\newcommand{\de}{\delta}
\newcommand{\pd}{\partial}
\newcommand{\Mr}{{\smM}}
\newcommand{\Md}{{\smM}_{\!D}}
\newcommand{\Mg}{{\smM}_{\!G}}
\newcommand{\Mq}{{\smM}_{\!{[q]}}}

\newcommand{\Mk}{{\smM}_{\!K}}
\newcommand{\Mp}{{\smM}_{\!Pl}}

\newcommand{\reg}{{\rm F}_{\{\Delg,\epsg\}}}
\newcommand{\sph}[1]{\Omega^{^{[#1]}}}

\newcommand {\be}{\begin{equation}} 
\newcommand{\fe}{\end{equation}}

\newcommand{\eqn}{\label}

\begin{document}
\title{Linearized self-forces for branes}
\author{Richard A. Battye}
\affiliation{Jodrell Bank Observatory, Department of Physics and
Astronomy, University of Manchester, Macclesfield, Cheshire SK11
9DL, UK}
\author{Brandon Carter}
\affiliation{LUTh, Observatoire de Paris--Meudon, 92195 Meudon, France}
\author{Andrew Mennim}
\affiliation{Institute of Cosmology and Gravitation, University of Portsmouth, Mercantile House, Hampshire Terrace, Portsmouth, PO1 2EG, UK}
\date{6 December 2004}

\begin{abstract}
We compute the regularized force density and renormalized action due
to fields of external origin coupled to  a brane of arbitrary
dimension in a spacetime of any dimension. Specifically, we consider
forces generated by gravitational, dilatonic and generalized
antisymmetric form-fields. The force density is regularized using
a recently developed gradient operator. For the case of a Nambu--Goto
brane, we show that the regularization leads to a renormalization of
the tension, which is seen to be the same in both approaches. We
discuss the specific couplings which lead to cancellation of the
self-force in this case.

\end{abstract}

\maketitle

\section{Introduction}
\label{intro}

This article reviews and extends the use of a convenient geometric method of
allowing for divergent self-interaction effects, generalizing to strings
and higher branes in arbitrary spacetime dimensions the kind of
regularization and renormalisation methods whose use for classical
point particles in four dimensions has long been familiar. This
problem was first discussed in the context of the point electron by
Dirac~\cite{dirac}, where it leads to the classical renormalization of
the electron mass, and has since been seen to be a more general
problem.

Our generalised method was originally developed for strings in an 
ordinary four dimensional background
in the context of electromagnetic effects~\cite{C97}, with
a view to its application to problems involving vortons~\cite{vorton}, and it
has since been extended to allow for the effect of linearized
gravitation\cite{BC95,CB98,C98b,C98a}.
For each of these interactions, it turns out that the divergent part
can be dealt with just by an appropriate renormalisation of the
worldsheet energy-momentum tensor, and of the relevant Lagrangian.
This applies in an extensive category of classical string
models~\cite{C89} for which in general the string tension
is less than the corresponding energy density, including the
kind~\cite{CP95} appropriate for describing the effects of 
Witten's superconductivity mechanism, and also the ``transonic''
kind~\cite{C95} describing the macroscopically averaged effects of
wiggles in a Nambu--Goto string.

An important special case is that of the Nambu--Goto (NG) model itself, for which
the string tension and the energy density are equal and constant, and for which it turns out
that, when only gravity is involved, the kind of renormalisation developed here
is not needed. In this particular  case, as discussed in detail in a
preceding article~\cite{CB98} the divergent contribution from
linearized self-force simply vanishes. This result is a prototype of
the cancellation theorem that has been extended to higher spacetime
dimensions~\cite{BD98a}, in which the gravitational divergence does
not cancel out by itself, but can be  cancelled by the axion and dilaton
contributions for standard values of the  relevant coupling constants
as obtained in the low energy limits of Superstring
theories~\cite{DH,gibbons,CHH90,C99}.

Recently, we have developed the mathematical machinery to apply this
formalism to higher-dimensional branes in a spacetime of arbitrary
dimension. In \cite{CBU} it was shown that the regularized gradient
operator applies more generally, and this was used in the case of
purely gravitational fields in \cite{BCM04}. The cancellation of the
gravitational self-force, already known for the case of a Nambu--Goto
string in four spacetime dimensions, was seen to hold for any
Nambu-Goto brane of co-dimension two, whatever the spacetime
dimension. It was pointed out that this could possibly provide a
mechanism for self-tuning of the cosmological constant in brane-world
models. 

In view of the considerable interest in brane models in the context
String/M-Theory both as low-energy limits and also via the fashionable
concept of brane-worlds, we have generalized the calculation to
include a range of coupled fields. The present discussion allows for
interactions mediated by long-range linearized dilaton and form-field
forces, as well as gravity. All specific cases discussed above are
seen to be special cases of this generalized calculation. We compute the
conditions on the couplings for the cancellation, already known
between the axion and dilaton in the case of a string in four
dimensions, in more general circumstances.

Before going into the technical details of the regularity problem that arises
in the case of self interaction, the first few sections will be 
devoted to the derivation of an appropriate formula for
the force density exerted by the linearized forces form-fields, the
dilaton and gravity on localised brane systems of a general kind in an
arbitrary number of dimensions.
It is necessary to go through this carefully because, although the
case of the form-field is a straight-forward generalization of the
Maxwell force, there are subtleties associated with the dilatonic and
gravitational forces.

\section{Brane worldsheet geometry}

In this section we shall explain our geometrical machinery
and notation.  Phrasing the problem in terms of this geometric
formalism allows results to be extended from specific solutions to general
configurations of the brane worldsheet in an elegant way which largely
avoids having to introduce new worldsheet coordinates.
This section is a recapitulation of the essential geometric concepts
needed for the kinematic description of the evolving worldsheet
as described for cosmic strings in~\cite{C92,C97a}.  We will be
considering a $p$-brane (with $(p+1)$-dimensional worldsheet) in an
$n$-dimensional ``bulk'' spacetime.

In this setup, the kinematics of the brane are described by the first
and second fundamental tensors of the worldsheet~\cite{C89,C97a}.
The first fundamental tensor $\etag^{\mu\nu}$ is simply the induced
metric on the worldsheet in terms of the bulk coordinates.
It can be written in terms of worldsheet position $\overline x{^\mu}$ and
worldsheet coordinates $\sigmag_i$ as
\be \etag^{\mu\nu}=\gamma^{ij}\frac{\pd\overline{x}^\mu}{\pd\sigmag^i}\,
\frac{\pd\overline{x}^\nu}{\pd\sigma^j}\,,\qquad
\gammag_{ij}=\gb_{\mu\nu}\frac{\pd\overline{x}^\mu}{\pd\sigma^i}
\frac{\pd\overline{x}^\nu}{\pd\sigma^j}\,,\eqn{7}\fe 
with $\gammag^{ij}$ being the components of the worldsheet metric
with respect to the worldsheet coordinates.
Lowering one index gives the projection operator onto the tangent
space of the brane, $\etag^\mu{}_\nu$.  One can also define an
orthogonal projection operator by
\be \perpg^{\!\mu}_{\,\nu}=\gb^\mu{_\nu}-\etag^\mu{_\nu}\,. \eqn{15}\fe
This allows us to give a covariant expression for the radial distance, $r$,
from the brane and the radially directed unit vector, $\nabla^\mu r$,
by
\begin{equation}
r^2=\perpg_{\mu\nu} x^\mu x^\nu\,,\qquad
\nabla^\mu r = r^{-1}\perpg^{\!\mu}_{\,\nu}x^\nu\,.
\end{equation}

The second fundamental tensor $\Kg_{\mu\nu}{^\rho}$, 
is given by
\be \Kg_{\mu\nu}{^\rho}=\etag^\sigma_{\ \nu}\overline\nabla_{\!\mu}
\etag^\rho_{\ \sigma}\, .\eqn{13}\fe
where the tangentially projected differentiation operator
$\overline\nabla_{\!\mu}$ is given by
\be \overline\nabla_{\!\mu} =
\etag_\mu^{\ \nu}\nabla_{\!\nu}\,.\eqn{8} \fe
The condition of integrability of the worldsheet is the Weingarten identity,
$\Kg_{[\mu\nu]}{^\rho}=0$, that is, that the second fundamental tensor is symmetric
under interchange of its first two indices~\footnote{In line with the
usual convention, we use round/square brackets to denote index
symmetrization/antisymmetrization}.
This tensor has the noteworthy property of being worldsheet orthogonal
on its last index, but tangential on either of the first two indices,
that is
\be \Kg_{\mu\nu}{^\sigma}\etag_\sigma{^\rho}=0
=\perpg^{\!\lambda}_{\,\mu} \Kg_{\lambda\nu}{^\rho}=0\,.\eqn{14}\fe
The only non-vanishing trace of this second fundamental tensor is the 
curvature vector
\be \Kg^\rho=\Kg_\mu{^{\mu\rho}}=\overline\nabla_{\!\nu}\etag^{\nu\rho}\, ,
\eqn{17}\fe 
which inherits worldsheet orthogonality, $\etag^\rho{_\sigma}\Kg^\sigma=0$.
This vector $\Kg^\rho$ is set to zero by the dynamical equations of motion 
in the special case of a Nambu--Goto string or brane model, as
discussed later, but not in more general models with non-trivial
internal structure.

We will be discussing a model with a dilaton, so it is essential to
distinguish the total Einstein metric
\be
\gb_{_{\rm E}\mu\nu}=\gb_{\mu\nu}+\hb_{\mu\nu}\,,\eqn{00}\, \fe
from the conformally related Dicke (or Jordan) metric given by 
$\gb_{_{\rm D}\mu\nu}=e^{2\phib}\, \gb_{_{\rm E}\mu\nu}$
where $\phib$ is the dilaton.
Our treatment is based on linearization of all the fields, meaning
that we will assume both $\phib$ and the metric perturbation
$\hb_{\mu\nu}$ to be small, so that the expression
\be \gb_{_{\rm D}\mu\nu}=\gb_{\mu\nu}+\hb_{\mu\nu}+2\phib\, \gb_{\mu\nu}
\,,\eqn{01}\fe
will be a sufficient approximation for the Dicke metric.

\section{Linearized long-range radiation fields}

The purpose of the present work is to deal with general brane models
of Nambu--Goto type, for which the tension ${\calT}$ of the brane is
equal to the energy per unit length ${\calU}$, and for which we
expect long-range gravitational and electromagnetic interactions to be
relevant.
Most of the previous literature on string self-interaction, as well as
being restricted to the Nambu--Goto case where ${\calT}= {\calU}$, has
considered only the effects of linearized dilatonic and axionic
couplings, which are physically more exotic, but technically simpler.
Effects of this last kind~\cite{VV87} are mediated by an antisymmetric
$q$-form gauge field, $\Br_{\mu_1\dots\mu_q}$, of the same general type that is
very useful in ordinary relativistic fluid mechanics~\cite{C94b}, with
a Wess--Zumino coupling.
It is possible to obtain the particular kind of fluid model appropriate
for the axion case in what is known as the Zel'dovich limit: the
``stiff'' limit characterised by the property that perturbations
propagate at the speed of light.
In the general fluid case, the sound speed can be lower, but this is
beyond the scope of our analysis.

Our work applies to fields satisfying the ordinary 
wave equation in the weak field limit, for which the background metric
$\gb_{\mu\nu}$ is Minkowski space.
These fields contribute to the $n$-dimensional action 
\begin{equation}
{\calI}_{\rm r}=\int\hat \Lr_{\rm r}
\Vert g\Vert^{1/2}\, {\rm d}^nx
\, ,\eqn{03a}
\end{equation}
that governs the behaviour of the long range radiation, where $\hat
\Lr_{\rm r}$ is the kinetic contribution the Lagrangian
density of the relevant fields.
It is helpful to introduce a factor of $\sph{n-2}$, where
$\sph{n}$ is defined as the surface area of a unit $n$-sphere, so as
to get the usual normalization of the Newton constant in the inverse-square law.
The values of $\sph{n}$ are given by
\begin{equation}
\sph{2j+1}=\frac{2\pi^{j+1}}{j!}\,, \qquad
\sph{2j}= \frac{2^{2j}\pi^j (j-1)!}{(2j-1)!}\,.
\end{equation}
For a given, $n$-dimensional background metric $\gb_{\mu\nu}$,
the radiation action density $\hat \Lr_{\rm r}$ will consist of
a sum of contributions that are homogeneously quadratic functions
of the gradient fields $\hb_{\mu\nu;\rho}$, $\phib_{;\rho}$
and $\Hr_{\mu_1...\mu_q}$ that are respectively
associated with the separate gravitational, dilatonic and $q$-form fields.
Explicitly, the relevant gravitational, dilatonic and $q$-form
field-strength tensors are given by 
\begin{equation}
\hb_{\mu\nu;\rho}= \nabla_{\!\rho}\hb_{\mu\nu}\, ,\qquad
\phib_{;\rho}=\nabla_{\!\rho}\phib\, ,\qquad
\Hr_{\mu_0\mu_1...\mu_q}=(q+1)\nabla\!_{[\mu_0}\Br_{\mu_1...\mu_q]}\,,
\end{equation}
where $\nabla_{\!\rho}$ is the operator of covariant differentiation as specified
with respect to the connection specified by the background metric
$g_{\mu\nu}$. The choice of connection evidently does not matter for the
definition of $\phi_{;\mu}$, nor, due to the antisymmetrisation, to that of
$\Hr_{\mu_1...\mu_q}$, but it does matter for that of the
gravitational field tensor $\hb_{\mu\nu;\rho}$.
It is to be understood that the same background metric
$\gb_{\mu\nu}$, rather than the associated Einstein metric $\gb_{_{\rm E}\mu\nu}$
or the associated Dicke metric $\gb_{_{\rm D}\mu\nu}$, is used
throughout for index raising and lowering.
The electromagnetic field is obviously an important special case
of the $q$-form field ($q=1$) so we need not include it explicitly.

We now consider the form of the part of the Lagrangian
$\hat \Lr_{\rm r}$ governing the long-range interactions.
If we use units with the speed of light $c$ and the Dirac constant
$\hbar$ set to unity, the long range radiation field contribution
action to the density can be expressed in terms of constant mass
parameters, $\Mg$, $\Md$ and $\Mq$, in the form
\begin{equation}
\hat \Lr_{\rm r}=\frac{1}{2\sph{n-2}}\Bigg\{
\frac{\Mg^{\, n-2}}{n-2}{\calR}^{(2)}_{_{\rm E}}
-\Md^{\, n-2}\phib^{;\mu}\phib_{;\mu}
-\frac{\Mq^{\, n-2-2q}}{(q+1)!}
\Hr^{\mu_0...\mu_q}\Hr_{\mu_0...\mu_q}\Bigg\}\,, \eqn{02}
\end{equation}
where ${\calR}^{(2)}_{_{\rm E}}$ is the residual
Ricci scalar contribution, as obtained from the measure weighted Ricci
scalar of the ordinary Hilbert action associated with the Einstein
metric (\ref{00}), by taking the expansion to quadratic order in the
metric perturbation $h_{\mu\nu}$ and ignoring divergences to remove
second-order derivatives. This leads
to the formula
\begin{equation}
{\calR}^{(2)}_{_{\rm E}}
=\frac{1}{2}\hb^{\mu\nu;\rho}\Bigg(\hb_{\rho\mu;\nu} -
\frac{1}{2}\hb_{\mu\nu;\rho} -\gb_{\mu\nu} \hb^{\rho\sigma}{_{;\sigma}}+
\frac{1}{2}\gb_{\mu\nu}\hb^\sigma{_{\sigma;\rho}}\Bigg)\, .\eqn{02g}
\end{equation}

The constant mass scales, $\Mg$, $\Md$ and $\Mq$, should not be
confused with the masses of the corresponding bosonic particles;
these bosons must have very small masses for the forces to qualify as
long-range and we will assume them to be massless in our treatment.
The mass scale $\Mg$ is the Planck mass of the
$n$-dimensional bulk spacetime, a term we will not use because of
possible confusion with the 4-dimensional Planck mass $\Mp=\hbox{\smG}^{-1/2}$,
where $\hbox{\smG}$ is Newton's constant. 
The mass scale of the coupling of the dilaton, $\Md$,
is usually supposed to be very large, at least comparable with the
gravitational mass.  In the usual case of ordinary spacetime ($n=4$),
solar system measurements provide severe observational
limits~\cite{D64} on the dimensionless Brans--Dicke parameter
$\omeg=2\hbox{\smG}\Md^{\,2}+3/2$, which must be very large
compared with unity, so that $\Md$ itself must large
compared with the Planck mass.
The other mass scale, $\Mq$, governs the coupling to the $q$-form field.
In the relevant 4-dimensional application to the axion field where $q=2$, 
the corresponding unrationalised pseudo-salar axion coupling constant,
as used by Battye and Shellard~\cite{BS95,BS96}, is related 
$M_{[2]}^2=2\pi\fr_{\rm a}^{\,2}$.  The value of this 
axion coupling mass scale is usually supposed to be considerably below the 
Planck mass. 

To obtain the field equations it is necessary to work out the
variational derivatives with respect to the fields on which the
action contributions depend.  These variational derivatives, the
vanishing of which is the condition for the field equations to be satisfied
in the source-free case, are given for the dilaton contribution by
\begin{equation}
\frac{\delta}{\delta\phib}\Big(-\phib^{;\rho}\phib_{;\rho}\Big)
=2\nabla_{\!\rho}\nabla^\rho\phib\,,
\end{equation}
and for the contribution from the $q$-form field by
\begin{equation}
\frac{\delta}{\delta \Br_{\mu_1...\mu_q}}\Big(-\Hr^{\rho_0...\rho_q}
\Hr_{\rho_0...\rho_q}\Big)=2(q+1)\nabla_{\!\rho}\Hr^{\rho\mu_1...\mu_q}\,.
\end{equation}
For the gravitational contribution, with a little more work and
using the vacuum property of the background metric, one 
obtains the formula
\begin{equation} \frac{\delta}{\delta \hb_{\mu\nu}}{\calR}^{(2)}_{_{\rm E}}
=\frac{1}{2}\nabla_{\!\rho}\nabla^{\rho}\hb^{\mu\nu}-\frac{1}{2}\gb^{\mu\nu}
\nabla_{\!\rho}\nabla^\rho\hb^{\mu\nu}-\nabla_{\!\rho}\nabla^{(\mu}\hb^{\nu)\rho}
+\frac{1}{2}\gb^{\rho(\mu}\nabla_{\!\rho}\nabla^{\nu)}\hb
+\frac{1}{2}\gb^{\mu\nu}\nabla_{\!\rho}\nabla_{\!\sigma}\hb^{\rho\sigma}\,.
\end{equation}

It is evident that both the action and the variational derivatives 
are unaffected by the gauge transformations
\begin{equation}
\Br_{\mu_1\mu_2...\mu_q}\mapsto \Br_{\mu_1\mu_2...\mu_q}+
q\nabla_{\![\mu_1} \chir_{\mu_2...\mu_q]}\,,
\end{equation}
for an arbitrary antisymmetric covector 
field $\chir_{\mu_2...\mu_q}$, and subject to the background metric satisfying
the Einstein vacuum field
equations, it can be checked that such
invariance also holds for gravitational gauge transformations of the form
\begin{equation}
\hb_{\mu\nu}\mapsto \hb_{\mu\nu}+ 2\nabla_{(\mu}\xib_{\nu)}\,,
\end{equation}
for an arbitrary displacement vector field $\xib^\mu$.

\section{Linearized interactions}

We will be considering situations where the linearized long-range
fields, whose kinetics are governed by the Lagrangian (\ref{02}), 
interact linearly with a material system described by a
Lagrangian contribution provided by a master function $\hat\Lambdar$,
that depends, possibly non-linearly, on a set of internal fields
representing physical quantities such as currents of various kinds, as
well as on the relevant background metric.
We shall consider only Lagrangians which do not depend on the
derivatives of the metric.

We are considering the linearized interaction, meaning that the
cross-coupling is governed by an interaction Lagrangian, $\hat \Lr_{\rm c}$, of the form
\be \hat \Lr_{\rm c}=\frac{1}{q!}\hat\Jr{^{\mu_1...\mu_q}}\Br_{\mu_1...\mu_q}
+\frac{1}{2}\hat \Tr{^{\mu\nu}}h_{\mu\nu} +\hat \Tr \phi\, ,\eqn{05}\fe
where the coefficients $\hat \Jr{^{\mu_1...\mu_q}}$, $\hat
\Tr{^{\mu\nu}}$, and $\hat \Tr$ are functions only of the internal field
quantities and the background metric that are involved in the specification
of the matter Lagrangian $\hat \Lambdar$.
We shall see that $\hat \Tr$ is indeed the trace of $\hat\Tr{^{\mu\nu}}$.
The total action of this underlying model takes the form
\be {\calI}=\int \hat \Lr\, \Vert g\Vert^{1/2}\, {\rm d}^{ n}x
=\int \left(\hat\Lambdar+\hat\Lr_{\rm r} + \hat\Lr_{\rm c}\right)
\, \Vert g\Vert^{1/2}\, {\rm d}^{ n}x\,.\eqn{06a}\fe

Although it was originally motivated by applications to the (most 
practically interesting) case of ordinary spacetime, for which $n=4$, 
the results developed here are particularly suitable for applications of 
a more speculative kind to cases for which $n$, the dimension of the
background spacetime, has higher values (including the values $n=10$
and $n=11$ that are of particular academic interest in the context of
Superstring theory and M-theory).
The ensuing field equations are obtained from the requirement 
of local invariance with respect to the variations of the linear interaction
fields $\Br_{\mu_1...\mu_q}$, $\hb_{\mu\nu}$ and $\phib$.
There are also various internal fields, which generally enter
non-linearly, involved in the specification of $\hat\Lambda$ and also
of the source coefficients $\hat\Jr{^{\mu_1...\mu_q}}$, $\hat
\Tr{^{\mu\nu}}$, and $\hat \Tr$.

The current coupling to the $q$-form field, $\hat
\Jr{^{\mu_1...\mu_q}}$, which is a generalization 
of an ordinary electric current vector or the vorticity flux bivector
of an axion, is restricted only by the condition that it must satisfy the flux conservation law,
\be \nabla_{\!\mu}\hat \Jr{^{\mu\mu_2...\mu_q}}=0\, ,\eqn{fluxcons} \fe
in order to be invariant under local Kalb--Ramond gauge transformations of the form
$\Br_{\mu_1\mu_2...\mu_q}\mapsto \Br_{\mu_1\mu_2...\mu_q}+q
\nabla_{\![\mu_1}\chir_{\mu_2...\mu_q]}$
for an arbitrary covector field $\chir_{\mu_2...\mu_q}$. 
For the special case of the electromagnetic field ($q=1$) this is
simply the current conservation condition, $\nabla_{\!\mu}\hat \Jr{^\mu}=0$,
necessary to ensure local gauge invariance under transformations of the form 
$\Br_\mu\mapsto \Br_\mu+\nabla_{\!\mu}\chir$ for an arbitrary scalar $\chir$.

Unlike the $q$-form field source term, the gravitational source term, $\hat \Tr{^{\mu\nu}}$, is
not something whose choice admits any latitude. In order for the theory under 
consideration to be considered as the linearization of a generally covariant 
model in which the gravitational field equations are obtained by requiring 
invariance with respect to variations of the total (Einstein) metric, it can 
be seen from the form of (\ref{00}) -- bearing in mind the metric dependence 
of the measure $\Vert \gb\Vert^{1/2}$ in (\ref{06a}) -- that this 
gravitational source term must be given by the geometric energy-momentum
tensor as defined by
\begin{equation}
\hat \Tr{^{\mu\nu}}= 2\Vert \gb\Vert^{-1/2} \frac{\partial \big(
\hat\Lambdar\Vert \gb\Vert^{1/2}\big)}{\partial \gb_{\mu\nu}}
= 2\frac{\partial \hat \Lambdar}{\partial \gb_{\mu\nu}}
+\hat \Lambdar \gb^{\mu\nu}\, .\eqn{07}
\end{equation} 
The scalar source coefficient $\hat \Tr$ might have various forms for
diverse scalar coupling theories that might be conceived.
However, in order for the coupling to be considered properly dilatonic it must be
derived from a model in which the original specification of the
matter Lagrangian was specified in terms of the
Dicke metric, as given in the linearized limit by (\ref{01}),
from which it can be seen that the resulting linearized coupling
coefficient in (\ref{05}) is necessarily the trace of $\Tr^{\mu\nu}$.
In the modern context of models derived in the low energy energy limit
from Superstring theory or M-theory the term dilatonic is often used for
theories involving scalar couplings of a more general kind
but, in most cases, such complications in the
underlying theory do not modify the form of the linearized limit to
which the present work is restricted.

Not much can be said about the equations of motion for the internal
fields characterising the material system until the form of its Lagrangian 
has been specified.
However, independently of such details, in a background spacetime of dimension $n$,
the equation of motion for the $q$-form field, using the
usual Lorentz gauge condition
\be \nabla^\mu \Br_{\mu\mu_2...\mu_q}=0\,, \fe
can be expressed in the usual~\cite{VV87} d'Alembertian form
\begin{equation}
\nabla_{\!\sigma}\nabla^\sigma
\Br_{\mu_1...\mu_q}=-\sph{n-2}\,\Mq^{\,2q+2-n}
 \hat \Jr_{\mu_1...\mu_q}\eqn{1b}\,.
\end{equation}
For the gravitational perturbation field $\hb_{\mu\nu}$ with a flat background
metric $\gb_{\mu\nu}$ is flat, using the de Donder gauge condition 
$2\nabla^\mu \hb_{\mu\nu}=\nabla_\nu \hb^\mu_{\,\mu}$,
the relevant source equation has the well known form
\begin{equation}
\nabla_{\!\sigma}\nabla^\sigma \hb_{\mu\nu}=
-2(n-2)\sph{n-2}\,\Mg^{\,2-n} \left(\hat \Tr_{\mu\nu}
-\frac{1}{n-2} \hat \Tr \gb_{\mu\nu}\right)\, , \eqn{2}
\end{equation}
where $n$ is the background  spacetime dimension. It is worth
remarking that, whereas, the pure gravitational radiation contribution is
strictly gauge invariant, the gravitational coupling contribution would be 
exactly gauge invariant only if $\nabla_{\!\mu} \hat \Tr^{\mu\nu}=0$,
a condition which would be satisfied only in the limit of
infinitely weak coupling.  This means that there will be a second order
discrepancy between this gravitational source equation (\ref{2}) and
the equation that would be obtained by rigorous application of the
formula (\ref{02g}).  Since we are working only to linear order, this
will not concern us.
For the dilaton field there is no similar issue of gauge and
\be
\nabla_{\!\sigma}\nabla^\sigma \phib=-\sph{\rm n-2}\, 
\Md^{\,2-n} \,\hat \Tr\,,\eqn{2b}\fe
is the relevant field equation.

\section{Distributional Sources}

There are problems inherent with modelling branes as distributional
sources, where the source densities $\hat \Jr{^{\mu_2...\mu_q}}$ and $\hat \Tr{^{\mu\nu}}$
are Dirac $\delta$-functions vanishing outside the worldsheet.
In the case where the codimension is greater than one this will
give rise to an ultra-violet (UV) divergence, as is most familiar in
the case $p=0$ of a point particle.
When one considers the full, non-linear Einstein equations, there are
further problems with distributional sources~\cite{traschen} which we
will not discuss here since we are only considering linearized
gravity.

In the case of a general $p$-brane, where the brane worldsheet has the
locus $x^\mu=\overline x{^\mu}\{\sigmag\}$ in terms of intrinsic
coordinates $\sigmag^\ig$ ($i=0,1, ..., p$), we can write the distributional
fields $\hat \Jr{^{\mu_1...\mu_q}}$ and $\hat\Tr{^{\mu\nu}}$ as
\begin{eqnarray}
\hat \Jr{^{\mu_1...\mu_q}}&=&\Vert \gb\Vert^{-1/2}\int \overline 
\Jr{^{\mu_1...\mu_q}}\, \delta^{[n]}[x-\overline x\{\sigmag\}]\, 
\Vert\gammag \Vert^{1/2}\, {\rm d}^{p+1}\sigmag\,,\eqn{3b}\\
\hat \Tr{^{\mu\nu}}&=&\Vert \gb\Vert^{-1/2}\int \overline \Tr{^{\mu\nu}}\, 
\delta^{[n]}[x-\overline x\{\sigmag\}]\, \Vert\gammag \Vert^{1/2}
\, {\rm d}^{ p+1}\sigmag\,,\eqn{4}
\end{eqnarray}
where $\Vert\gammag\Vert$ is the determinant of the induced metric.
The field $\overline \Jr{^{\mu_1...\mu_q}}$ is the generalised surface
current of the $q$-form field and is a regular vector field on
the brane worldsheet, but undefined off it.
The flux conservation law (\ref{fluxcons}) will also apply to this
surface current.
We define $\overline \Tr{^{\mu\nu}}$ similarly.
The same is true of the Lagrangian, which can be written
\be \hat \Lr=\Vert \gb\Vert^{-1/2}\int \overline \Lr\, 
\delta^{\rm n}[x-\overline x\{\sigmag\}]\, \Vert\gammag \Vert^{1/2}
\, {\rm d}^{p+1}\sigmag\, ,\fe
and similarly for the component contributions $\overline\Lambdar$
and $\overline \Lr_{_{\rm c}}$.
The master function, $\overline\Lambdar$, will be the intrinsic worldsheet 
Lagrangian, which is a function just of the relevant internal
fields, such as currents, on the string, and of its induced metric,
while the cross coupling contribution, $\overline \Lr_{_{\rm c}}$,
will be given in terms of the worldsheet confined fields $\overline 
\Jr{^{\mu_1...\mu_q}}$ and $\overline \Tr{^{\mu\nu}}$ by 
\be \overline \Lr_{_{\rm c}}=
\frac{1}{q!}\overline \Jr{^{\mu_1...\mu_q}}\Br_{\mu_1...\mu_q}
+\frac{1}{2}\overline \Tr{^{\mu\nu}}\hb_{\mu\nu}
+\overline \Tr \phib\, .\eqn{LcBar} \fe
The action can then be expressed, without any distributional terms, as a simple
$(p+1)$-surface integral
\be {\calI}=\int \overline \Lr\, \Vert \gammag\Vert^{1/2}\,  
{\rm d}^{p+1}\sigmag\, .\eqn{56a}\fe

The surface energy-momentum tensor $\overline \Tr{^{\mu\nu}}$, which
appears in $\overline \Lr_{_{\rm c}}$, can be obtained directly from
the worldsheet master function, $\overline\Lambdar$, without the use
of distributions, by variation with respect to the metric, giving the
formula
\be \overline \Tr{^{\mu\nu}}= 2\Vert\gammag\Vert^{-1/2}\frac{\partial
(\overline\Lambdar\Vert\gammag\Vert^{1/2})}{\partial \gb_{\mu\nu}}\,.\eqn{58}\fe
To vary the part of the action coming from $\overline \Lr_{_{\rm c}}$, we need the 
variational derivative of $\overline \Tr{^{\mu\nu}}$ with respect to
the metric.  This motivates us to define what we call the
\emph{hyper-Cauchy tensor} (a relativistic generalisation of the
Cauchy elasticity tensor of classical mechanics) which is defined by
\begin{equation}
\overline{\calC}^{\mu\nu\rho\sigma}=\Vert\gammag\Vert^{-1/2}
\frac{\de}{\de\gb_{\mu\nu}}\Big(\overline \Tr{^{\rho\sigma}}
\Vert\gammag\Vert^{1/2}\Big)=2\Vert\gammag\Vert^{-1/2}
\frac{\de}{\de \gb_{\mu\nu}}\frac{\de}{\de\gb_{\rho\sigma}}
\Big(\overline\Lambdar\Vert\gammag\Vert^{1/2}\Big)
=\overline{\calC}{^{\rho\sigma\mu\nu}} \,, \eqn{59}
\end{equation}
and is manifestly symmetric under the interchange of the first pair of
indices with the second pair.
We can rewrite the expressions (\ref{58}) and (\ref{59}) in the more
practical forms
\begin{equation}
\overline \Tr{^{\mu\nu}}= 
2\frac{\partial\overline\Lambdar}{\partial \gb_{\mu\nu}}
+\overline\Lambdar \etag^{\mu\nu}\,,\qquad
\overline{\calC}{^{\mu\nu\rho\sigma}}=\frac{\delta \overline 
\Tr{^{\rho\sigma}}}{\delta \gb_{\mu\nu}}+\frac{1}{2}
\overline \Tr{^{\rho\sigma}} \etag^{\mu\nu}\,,\eqn{59b}
\end{equation}
and both the energy-momentum tensor and the hyper-Cauchy tensor are
tangent to the brane, that is,
\begin{equation}
\perpg^\lambda{}_\mu\overline \Tr{^{\mu\nu}}=
\perpg^\lambda{}_\mu\overline{\calC}{^{\mu\nu\rho\sigma}}=0\,.
\end{equation}
The generalized current $\overline\Jr{^{\mu_1...\mu_q}}$ is
a conserved flux, so its variation with respect to the metric is a
total derivative, which will vanish in the action integral.

If the corresponding radiation fields $\Br_{\mu_1...\mu_q}$, 
$\hb_{\mu\nu}$ and $\phib$ are considered to be regular background fields
due to external sources, the treatment of such a system will
be straightforward, but it is evident that this will not be the case
for the radiation fields produced by the brane itself, since they
will be singular at the brane just where their evaluation is needed.

The introduction of these fields enables the maximally symmetric (static,
asymptotically vanishing) solution of the simple wave equation
(\ref{2b}) for a uniform  $p$-brane supported distribution of
the form (\ref{4}) to be expressed in the form
\begin{equation}
\phib^{;\rho}=-\frac{\sph{n-2}}{\sph{n-2-p}}\,
\Md^{\,2-n} \,\overline \Tr\, \perpg^{\!\rho}_{\,\sigma}
\frac{x^\sigma}{r^{n-1-p}}\,,
\end{equation}
in which the angle factor $\sph{n-2}/\sph{n-2-p}$
evidently reduces to unity in the case of a point particle ($p=0$).
The analogous expressions for the form and gravitational fields are given by
\begin{equation}
\Br^{\mu_1...\mu_q;\rho}=-\frac{\sph{n-2}}{\sph{n-2-p}}\,
\Mq^{\,2q+2-n}\,\overline \Jr{^{\mu_1...\mu_q}}\,
 \perpg^{\!\rho}_{\,\sigma}\frac{x^\sigma}{r^{n-1-p}}\,,
\end{equation}
and
\begin{equation}
\hb^{\mu\nu;\rho}=-2(n-2)\frac{\sph{n-2}}{\sph{n-2-p}}\,
\Mg^{\,2-n}\,\Big(\overline \Tr^{\mu\nu}
-\frac{1}{n-2} \overline \Tr \gb^{\mu\nu}\Big)
\perpg^{\!\rho}_{\,\sigma}\frac{x^\sigma}{r^{n-1-p}}\,.
\end{equation}

\section{The force density formulae}

To derive the equations of motion from a variation principle, we must consider
perturbative displacements with respect to the background
characterised by the metric $\gb_{\mu\nu}$ and the linearly coupled
$\hb_{\mu\nu}$, $\phib$ and $\Hr_{\mu_1\cdots\mu_q}$ fields.
We find it most convenient to describe the effect of displacements
using a Lagrangian treatment where the background coordinates $x^\mu$
are considered to be dragged along by the displacement, so that the
relevant field variations are given just by the corresponding Lie
derivatives with respect to the vector field $\xib^\mu$ describing the
displacement under consideration.
This leads to the formulae
\begin{eqnarray}
\delta \Br_{\mu_1\mu_2...\mu_q}&=&\xib^\sigma\nabla_{\!\sigma} 
\Br_{\mu_1\mu_2...\mu_q}+q \Br_{\sigma[\mu_2...\mu_q}
\nab_{\!\mu_1]}\xib^\sigma \,,\eqn{70b} \\
\delta \hb_{\mu\nu}&=&\xib^\sigma\nab_{\!\sigma}\hb_{\mu\nu}+
2 \hb_{\sigma(\mu}\nab_{\!\nu)}\xib^\sigma \eqn{70c} \,,\\
\delta \phib&=&\xib^\sigma\nab_{\!\sigma}\phib \eqn{70d} \,,
\end{eqnarray}
for the $q$-form, gravitational and dilatonic fields respectively,
while finally for the background metric itself one has the well known
formula
\be \delta \gb_{\mu\nu}=2\nab_{\!(\mu}\xib_{\nu)}\, .\eqn{70e} \fe

There are, of course, the internal fields on which the master function
$\Lambdar$ depends and these must also be perturbed in a full
variational analysis.
However, if the internal field equations are satisfied, these
perturbations will have no effect on the action integral ${\calI}$ so,
for the purpose of evaluating the variation $\delta{\calI}$, there
will be no loss of generality in assuming that these fields are unperturbed.
The worldsheet flux conservation law (\ref{fluxcons}) tells us~\cite{C92} that
$\overline\Jr{^{\mu_1...\mu_q}}$ is related by Hodge duality to the
exterior derivatives of corresponding worldsheet $(p-q+1)$-forms.
In the usual cases these differential forms will be included among (or
depend only on) the relevant internal fields whose variation we can
legitimately ignore for  the purpose of evaluating $\delta{\calI}$, so
the variation of these $(p-q+1)$-forms can also be taken to be zero.
This means that the variations of the corresponding surface density
will also vanish, that is,
\be
\delta\Big(\Vert\gammag\Vert^{1/2} \overline \Jr{^{\mu_1...\mu_q}}\Big)=0\, 
. \eqn{72b} \fe
It follows that the contribution from the $q$-form field to the
variation of (\ref{LcBar})  will be given by 
\be \delta\Big(\Vert\gammag\Vert^{1/2}\overline \Jr{^{\mu_1...\mu_q}}
\Br_{\mu_1...\mu_q}\Big)=\Vert\gammag\Vert^{1/2}\overline \Jr{^{\mu_1...\mu_q}}\,
\delta \Br_{\mu_1...\mu_q} \,.\eqn{73a}\fe

The variation of the background metric does not contribute to
the variation of the $q$-form terms in (\ref{LcBar}), but it is of
paramount importance for the evaluation of the corresponding
contribution from the gravitational and dilatonic coupling
terms.
It can be seen from (\ref{59b}) that the gravitational contribution to the 
variation of the integrand in (\ref{56a}) will be given by the -- until
recently~\cite{BC95} not so well known -- expression
 \be \delta\Big(\frac{1}{2}\Vert\gammag\Vert^{1/2}\ \overline \Tr{^{\mu\nu}} 
\hb_{\mu\nu}\Big)=\frac{1}{2}\Vert\gammag\Vert^{1/2}\Big(
\overline \Tr{^{\mu\nu}}\delta \hb_{\mu\nu} +\overline{\calC}
{^{\mu\nu\rho\sigma}}\hb_{\rho\sigma}\delta \gb_{\mu\nu}\Big)\, ,\eqn{73c}\fe
while, despite its deceptively simple scalar nature, the dilatonic 
coupling gives rise to a corresponding contribution that works out to be
given by  the  -- even less well known -- expression
 \be \delta\Big(\Vert\gammag\Vert^{1/2}\ \overline \Tr \phib\Big)= 
\Vert\gammag\Vert^{1/2}\Big( \overline \Tr \delta\phib + 
(\overline \Tr{^{\mu\nu}}+ \overline{\calC}{^{\mu\nu}} )\phib
 \,\delta \gb_{\mu\nu}\Big)\, ,\eqn{73d}\fe
using the notation $\overline{\calC}{^{\mu\nu}} = \overline{\calC}{^{\mu\nu\rho}}_\rho$.

When the internal field equations are satisfied, the variation of the
background metric will of course provide the only contribution from
the term involving the master function $\overline\Lambdar$ in the
Lagrangian.
As can be seen from (\ref{58}), this last contribution will simply be
given by an expression of the familiar form
\be  \delta\Big(\Vert\gammag\Vert^{1/2}\Lambdar\Big)=
\frac{1}{2}\Vert\gammag\Vert^{1/2}\,
\overline \Tr{^{\mu\nu}}\delta \gb_{\mu\nu}\, .\eqn{73}\fe

To evaluate the integrated effect of these contributions
(\ref{73a}--\ref{73}) we substitute the relevant Lie derivative formulae
(\ref{70b}--\ref{70e}) and write the terms involving derivatives of
the displacement fields as total divergences. Using the current
conservation law (\ref{fluxcons}) one finds
\begin{equation}
\frac{1}{q!}\overline \Jr{^{\mu_1...\mu_q}}\delta\,\Br_{\mu_1...\mu_q}=
\frac{1}{q!} \xib^\mu \Hr_{\mu\mu_1...\mu_q} \overline \Jr{^{\mu_1...\mu_q}} 
+\frac{1}{(q-1)!}\overline \nabla_{\!\mu}\big(\xib^\nu \Br_{\nu\mu_2...\mu_q}
\overline \Jr{^{\mu\mu_2...\mu_q}}\big)\, ,\eqn{74b}
\end{equation}
from (\ref{73a}). For the first term in (\ref{73c})
\begin{equation}
\frac{1}{2}\overline \Tr{^{\mu\nu}}\delta \hb_{\mu\nu}
=\xib^\mu\left[\frac{1}{2}\overline \Tr{^{\nu\rho}}\nab_{\!\mu} \hb_{\nu\rho}
-\overline\nabla_{\!\nu}\Big(\overline \Tr{^{\nu\rho}}\hb_{\mu\rho}\Big)\right]
+\overline\nabla_{\!\mu}\Big(\xib^\nu\overline \Tr{^{\mu\rho}\hb_{\nu\rho}}
\Big)\, ,\eqn{74c}
\end{equation}
and for the second term in (\ref{73c}) 
\begin{equation}
\frac{1}{2}\overline{\calC}{^{\mu\nu\rho\sigma}}\hb_{\rho\sigma}
\delta \gb_{\mu\nu}=-\xib^\mu\overline\nabla_{\!\nu}\Big(
\overline{\calC}{^\nu}_\mu{^{\rho\sigma}}\hb_{\rho\sigma}\Big)+
\overline\nabla_{\!\mu}\Big(\xib^\nu
\overline{\calC}{^\mu}_\nu{^{\rho\sigma}}\hb_{\rho\sigma}\Big)\,
.\eqn{74d}\end{equation}
The corresponding expression for the second term in (\ref{73d}) is
\begin{equation}
\big(\overline \Tr{^{\mu\nu}}+\overline{\calC}{^{\mu\nu}}\big)\phib\,
\delta \gb_{\mu\nu}=-2\xib^\mu\overline\nabla_{\!\nu}\Big[\big(
\overline \Tr{^\nu}_\mu+ \overline{\calC}{^\nu}_\mu\big)\phib\Big]+
2\overline\nabla_{\!\mu}\Big[\xib^\nu\big(\overline \Tr{^\mu}_\nu+
\overline{\calC}{^\mu}_\nu\big)\phib \Big]\, ,\eqn{74e}
\end{equation}
while for the first term in (\ref{73d}) one trivially obtains
\be\overline \Tr\delta\phib=\xib^\mu\overline \Tr\nabla_{\!\mu}\phib
\, .\eqn{74f}\fe
Finally, for the intrinsic contribution given by (\ref{73}) 
one obtains an expression of the familiar form
\begin{equation}
\frac{1}{2}\overline \Tr{^{\mu\nu}}\delta \gb_{\mu\nu} =-\xib^\mu
\overline\nabla_{\!\nu}\overline \Tr{^\nu}_\mu+\overline\nabla_{\!\mu}
\big(\xib^\nu \overline \Tr{^\mu}_\nu\big)\, .\eqn{74g}
\end{equation}

To derive the force density formulae, we vary the action integral
(\ref{56a}) with respect to $\xib_\mu$, the displacement.
We wish apply Green's theorem to remove the divergence terms, so we must
require that the displacement be confined to a finite region.
The result is an expression of the form
\be \delta {\calI}=\int \xib^\mu\big( \overline \fr_\mu
-\overline\nabla_{\!\nu}\overline \Tr{^\nu}_\mu 
\big) \Vert \gammag\Vert^{1/2}\, {\rm d}^{p+1}\sigmag\, ,
\fe
so that applying the variation principle gives the equation of motion
\be \overline\nabla_{\!\nu}\overline \Tr{^{\mu\nu}}= \overline \fr{^\mu}
\eqn{20} \fe
in which the vector $\overline \fr{^\mu}$ represents the total force density 
exerted by the various radiation fields involved. This force density can immediately be read out in the form
\be \overline \fr{^\mu}=\overline\fr_{_{\![q]}}{^\mu}+
\overline \fr_{_{\!\rm G}}{^\mu}+\overline\fr_{_{\!\rm D}}{^\mu}\, ,\fe
in which the contributions from the various fields involved are as follows.
The $q$-form contribution can be seen from (\ref{74b}) to be given by
\begin{equation}
\overline \fr_{_{\![q]}}{^\mu}=\frac{1}{q!}\Hr^\mu{_{\mu_1...\mu_q}}
\overline \Jr{^{\mu_1...\mu_q}} \eqn{21b}\,,
\end{equation}
which, in the case $q=1$, is the Lorentz force of electromagnetism.
The gravitational contribution can be seen from (\ref{74c}) and (\ref{74d}) 
to be given by
\be \overline \fr_{_{\!\rm G}}{^\mu}=\frac{1}{2}\overline \Tr{^{\nu\sigma}}
\nab^\mu \hb_{\nu\sigma} - \overline\nabla_{\!\nu}\big( \overline 
\Tr{^{\nu\sigma}} \hb_\sigma{^\mu}+ \overline{\calC}{^{\mu\nu\rho\sigma}} 
\hb_{\rho\sigma} \big) \, . \eqn{22a}\fe
Finally, the dilatonic contribution can be seen from (\ref{74e}) and 
(\ref{74f}) to be given by the expression
\begin{equation}
\overline \fr_{_{\!\rm D}}{^\mu} =\overline \Tr\nabla^\mu\phib
-2\overline\nabla_{\!\nu}\Big[\big(\overline \Tr{^{\mu\nu}}+
\overline{\calC}{^{\mu\nu}}\big)\phib\Big] \,.\eqn{22b}
\end{equation}
It has been shown~\cite{BC95} that some of
the early work on
cosmic strings~\cite{CHH90} is flawed due to omission of some of the terms and,
consequently, the related physical effects.

If the system of equations governing the dynamics of the internal
fields in the brane worldsheet involves $Q$ independent
degrees of freedom, the complete set of dynamical equations governing 
the evolution of the brane will involve a total of $Q+p+1$ degrees of
freedom, including those needed to determine the geometrical
evolution of the $p+1$ dimensional supporting worldsheet.
Since it involves $n$ components, the force law (\ref{20}) will, by itself,
 provide a complete system of equations of motion for the system if $Q<n-p$.
There will even be some redundancy, in the sense that the equations of 
the system (\ref{20}) will not all be mutually independent,
in cases for which $Q<n-p-1$. In the case of a
brane of the simple Nambu--Goto type for which $Q=0$ because there are
no internal fields, this is particularly so.

\section{Allowance for regularized self interaction}

Before proceeding further, it is convenient to decompose the various
linear perturbation fields under consideration in the form 
\begin{equation}
\Br_{\mu_1\dots\mu_q}=\widetilde \Br_{\mu_1...\mu_q}+\widehat \Br_{\mu_1...\mu_q}
\,,\qquad \hb_{\mu\nu}=\widetilde \hb_{\mu\nu}+\widehat \hb_{\mu\nu}\,,\qquad
\phib=\widetilde \phib+\widehat \phib\, ,\eqn{28}
\end{equation}
using a tilde for the locally source free contributions
$\widetilde \Br_{\mu_1...\mu_q}$, $\widetilde \hb_{\mu\nu}$
and $\widetilde \phib$,  respectively attributable to incident
$q$-form, gravitational and dilatonic radiation,
and a hat for the contributions given by the retarded Green 
function solutions of the relevant source equations (\ref{1b}), 
(\ref{2}) and (\ref{2b}). This will give rise to corresponding 
decompositions 
\begin{equation}
{\bar\fr}_{_{\![q]}}^{\,\mu}=\widetilde \fr_{_{\![q]}}^{\,\mu}
+\widehat \fr_{_{\![q]}}^{\,\mu}\,,
\qquad {\bar\fr}_{_{\!\rm G}}^{\,\mu}=\widetilde \fr_{_{\!\rm G}}^{\,\mu}
+\widehat \fr_{_{\!\rm G}}^{\,\mu} \,\qquad
{\bar\fr}_{_{\!\rm D}}^{\,\mu}=\widetilde \fr_{_{\!\rm D}}^{\,\mu}
+\widehat \fr_{_{\!\rm D}}^{\,\mu}\,, \eqn{30}
\end{equation}
for the associated force densities as specified by the general formulae 
(\ref{21b}), (\ref{22a}), (\ref{22b}).

In many contexts the coupling is so weak that the self-force contributions
$\widehat \fr_{_{\![q]}}^{\,\mu}$, $\widehat \fr_{_{\!\rm G}}^{\,\mu}$
and $\widehat \fr_{_{\!\rm D}}^{\,\mu}$ can be neglected.
However, in cases for which one needs to take account of the self
induced contributions $\widehat \Br_{\mu_1...\mu_q}$, $\widehat
\hb_{\mu\nu}$ and $\widehat \phib$,
one runs into difficulties arising from the fact that the field
equations (\ref{1b}), (\ref{2}) and (\ref{2b}) will be sourced by
the distributional fields, $\hat\Jr{^{\mu_1...\mu_q}}$, $\hat
\Tr{^{\mu\nu}}$ and $\hat \Tr$, rather than the regular worldsheet
supported fields $\overline \Jr{^{\mu_1\dots\mu_q}}$, and
$\overline \Tr{^{\mu\nu}}$. 
For sources such as these, the resulting field contributions
will diverge in the thin worldsheet limit if $n-p>1$.

To regularize this, we observe that the infinitely thin worldsheet is
an approximation of the physical object, which has finite thickness, $\epsg$,
giving an ultraviolet (UV) cut-off scale.
This will be sufficient to regularize point particles and extended
objects of codimension greater than two, that is, $n-p>3$, in which cases
the divergence will be of power law type. However, in the
logarithmically divergent case of a hyperstring of codimension
2, it will also be necessary to introduce a long-range infrared (IR)
cut off length scale, $\Delg$, that might represent the macroscopic
mean distance between neighbouring hyperstrings or the
compactification radius of extra-dimension in a brane-world model.  

We can deal with all these cases, leaving aside only the
hypersurface case of codimension one, by introducing a regularization
factor~\cite{CBU,BCM04} of the form
\begin{equation}
\reg=\frac{\sph{n-2}\sph{p}}{\sph{n-1}}\int_{\epsg}^{\Delg} x^{p-n+2}\,{\rm d}x\,,
\end{equation}
which should be accurate upto a factor of ${\cal O}(1)$.
This will be proportional to $\epsg^{p-n+3}$ (assuming $\Delg$ to be
large) when $p+3<n$ and $\log(\Delg/\epsg)$ when $p+3=n$.

It can then be seen from (\ref{1b}) that the regularized $q$-form self
field contribution will be given by
\begin{equation}
\widehat \Br_{\mu_1...\mu_q}=\frac{1}{n-2}\reg\,\Mq^{\,2q+2-n}\,
\overline\Jr_{\mu_1...\mu_q}\,  .\eqn{32b}
\end{equation}
Similarly, from (\ref{2}) the corresponding expression 
for the regularized gravitational self-field, $\widehat \hb_{\mu\nu}$, 
will be
\begin{equation}
\widehat \hb_{\mu\nu}=2\,\reg\,\Mg^{\,2-n}\,
\big(\overline \Tr_{\!\mu\nu}-\frac{1}{n-2}
\overline \Tr_{\!\sigma}{^\sigma}\gb_{\mu\nu}\big)\, ,\eqn{33}
\end{equation}
while finally by (\ref{2b}) we find that
\begin{equation}
\widehat\phib=\frac{1}{n-2}\reg\,\Md^{\,2-n}\,\overline \Tr\,,\eqn{33b}
\end{equation}
for the regularized dilatonic self-field.
For most purposes it will be adequate to use the 
same regularization factor $\reg$ for all the different fields.
This is especially true in the case the hyperstring case, where the 
dependence on the cut-off is only logarithmic.

In order to obtain correspondingly regularized self-force
contributions $\widehat \fr_{_{\![q]}}^{\,\mu}$, $\widehat
\fr_{_{\!\rm G}}^{\,\mu}$ and $\widehat \fr_{_{\!\rm D}}^{\,\mu}$ from
the formulae (\ref{21b}), (\ref{22a}) and (\ref{22b}), we need to know
not only the regularized values of the self-fields $\widehat
\Br_{\mu_1...\mu_q}$, $\widehat \hb_{\mu\nu}$ and $\widehat\phib$, but
also the regularized values of their gradients.
There is no difficulty for the terms involving just the tangentially
projected gradient operator $\overline\nabla_{\!\nu}$ but there are
also contributions from the unprojected gradient operator
$\nab_{\!\nu}$ which is meaningful only when acting fields whose
support extends off the worldsheet.  

Fortunately, this problem has a very simple general
solution~\cite{CBU}, of which particular applications in particular
gauges are implicit in  much previous work 
\cite{CHH90,BS95,BS96,BD98b} and which was first
formulated explicitly in the specific context of the electromagnetic
force in the string case~\cite{C97}. One finds, by examining
the string worldsheet limit behaviour of derivatives of the relevant
Green function, that the appropriate regularization of the
gradients on the string worldsheet is obtained simply by replacing the
ill-defined $\nab_{\!\nu}$ by the corresponding regularized gradient
operator given in terms of the worldsheet curvature vector $\Kg^\mu$
by the formula 
\be\widehat\nabla_{\!\nu}=\overline\nabla_{\!\nu}+\frac{1}{2}\Kg_\nu \,
.\eqn{grad}\fe 


Applying (\ref{grad}) to the $q$-form force contribution in (\ref{21b}) one finds
that it can be formulated as a worldsheet divergence and written in the form
\be\widehat \fr_{_{\![q]}}^{\,\mu}=-\overline\nabla_{\!\nu}
\widehat \Tr_{_{\![q]}}{^{\mu\nu}} \, ,\eqn{37a}\fe
in which
\be \widehat \Tr_{_{\![q]}}{^{\mu\nu}}=\frac{1}{(q-1)!}\widehat 
\Br{^\mu}_{\rho_2...\rho_q} \overline \Jr{^{\nu\rho_2...\rho_q}} 
-\frac{1}{2q!}\widehat \Br_{\rho_1..\rho_q}
\overline \Jr{^{\rho_1...\rho_q}}\etag^{\mu\nu} \, .\eqn{38a}\fe
When one applies same procedure to the gravitational self-force contribution 
in (\ref{22a}) one finds~\cite{C98a} that it too can be formulated as a worldsheet divergence in the analogous form
\be\widehat \fr_{_{\!\rm G}}^{\,\mu}=-\overline\nabla_{\!\nu}
\widehat \Tr_{_{\!\rm G}}{^{\mu\nu}}
\, ,\eqn{37c}\fe
in which the relevant energy-momentum contribution from the
gravitational self-interaction works out to be given by the expression
\be\widehat \Tr_{_{\!\rm G}}{^{\mu\nu}}=\widehat \hb_\sigma{^\mu}
\overline \Tr{^{\nu\sigma}}-\frac{1}{4}\widehat \hb_{\rho\sigma}
\overline \Tr{^{\rho\sigma}}\etag^{\mu\nu}+\widehat \hb_{\rho\sigma}
\overline{\calC}{^{\rho\sigma\mu\nu}}\, .\eqn{38c}\fe
Similarly  for the dilatonic contribution in (\ref{22b}) one obtains
\be\widehat \fr_{_{\!\rm D}}^{\,\mu}=-\overline\nabla_{\!\nu}
\widehat \Tr_{_{\!\rm D}}{^{\mu\nu}} \, ,\eqn{37d}\fe
with
\be\widehat \Tr_{_{\!\rm D}}{^{\mu\nu}}=2\widehat\phib\big(\overline 
\Tr{^{\mu\nu}}-\frac{1}{4}\overline \Tr\etag^{\mu\nu}+\overline
{\calC}{^{\mu\nu}}\big) \, .\eqn{38d}\fe

The remarkable fact that such a formulation exists is what makes it possible 
to describe the the result of this regularization as a ``renormalisation'':  
the possibility of expressing the self-force contributions  as divergences  
implies that that they can be absorbed into the left hand side of the basic 
force balance equation by a renormalisation whereby the original ``bare''
energy-momentum tensor $\overline \Tr{^{\mu\nu}}$ undergoes a 
replacement $\overline \Tr{^{\mu\nu}}\mapsto\widetilde\Tr{^{\mu\nu}}= 
\overline \Tr{^{\mu\nu}}+\widehat \Tr{^{\mu\nu}}$
with
\begin{eqnarray} \widehat \Tr{^{\mu\nu}}=\widehat \Tr_{_{\![q]}}{^{\mu\nu}}+
\widehat \Tr_{_{\!\rm G}}{^{\mu\nu}}+\widehat \Tr_{_{\!\rm D}}{^{\mu\nu}}
&=&\widehat \hb_\sigma{^\mu}
\overline \Tr{^{\nu\sigma}}-\frac{1}{4}\widehat \hb_{\rho\sigma}
\overline \Tr{^{\rho\sigma}}\etag^{\mu\nu}+\widehat \hb_{\rho\sigma}
\overline{\calC}{^{\rho\sigma\mu\nu}}+
2\widehat\phib\big(\overline 
\Tr{^{\mu\nu}}-\frac{1}{4}\overline \Tr\etag^{\mu\nu}+\overline
{\calC}{^{\mu\nu}}\big)\nonumber\\&&+
\frac{1}{(q-1)!}\widehat 
\Br{^\mu}_{\rho_2...\rho_q} \overline \Jr{^{\nu\rho_2...\rho_q}} 
-\frac{1}{2q!}\widehat \Br_{\rho_1..\rho_q}
\overline \Jr{^{\rho_1...\rho_q}}\etag^{\mu\nu}\,.
\label{newT}
\end{eqnarray}

The force balance equation (\ref{20}) can thereby be rewritten as
\be\overline\nabla_{\!\nu}\widetilde \Tr{^{\mu\nu}}= \widetilde \fr {^\mu}
\, ,\fe
in which the force density terms on the right consist just of well behaved 
locally source-free contributions from any incident radiation,
as given by the sum
\be \widetilde \fr {^\mu}=
\widetilde \fr_{_{\![q]}}{^\mu}+\widetilde \fr_{_{\!\rm G}}{^\mu}+
\widetilde \fr_{_{\!\rm D}}{^\mu}
\, ,\eqn{44}\fe
in which each of the terms is entirely regular.

\section{Action renormalisation}

It has just been demonstrated that the dominant contributions to the
$q$-form, gravitational and dilatonic self-interactions, can be
described in terms a renormalised energy-momentum tensor.
We now show that this renormalised energy-momentum tensor can be
derived by variational methods from a renormalized
action, in which the original Lagrangian master function, $\Lambdar$, is 
replaced by an renormalised function, $\widetilde\Lambdar$.

In order to incorporate the effects of self-interaction, as described by
the renormalised force balance equation (\ref{44}), it can be verified that 
all one needs to do is to replace the Lagrangian ${\calL}$ by
\begin{equation}
\widetilde{\calL}= \widehat\Lambdar
+\frac{1}{2(q!)} \widetilde \Br_{\mu_1...\mu_q} \overline 
\Jr{^{\mu_1...\mu_q}}+\frac{1}{4} \widetilde \hb_{\mu\nu}\overline
\Tr{^{\mu\nu}}+\frac{1}{2}\widetilde \phib\overline \Tr  \,, \eqn{62}
\end{equation}
where the last three terms are the effects from non-local
contributions to the fields originating far away from the brane,
and the renormalised master function is $\widetilde\Lambdar=\widehat\Lambdar+
\widehat\Lambdar_{_{[q]}} + \widehat\Lambdar_{_{\rm G}} + 
\widehat\Lambdar_{_{\rm D}}$, with renormalization terms added for
each of the fields.
The $q$-form contribution will be given by the expression
\begin{equation}
\widehat\Lambdar_{_{[q]}}=\frac{1}{2(q!)}\widehat \Br_{\mu_1...\mu_q}
\overline \Jr{^{\mu_1...\mu_q}}=\frac{1}{2(q!)(n-2)}\reg\,\Mq^{\,2q+2-n}
\,\overline \Jr_{\mu_1...\mu_q}\overline 
\Jr{^{\mu_1...\mu_q}} \,.\eqn{Lambdaq}
\end{equation}
The corresponding gravitational contribution has been evaluated in
 ~\cite{C98b} and has the form
\begin{equation}
\widehat\Lambdar_{_{\rm G}}=\frac{1}{4}\widehat \hb_{\mu\nu} 
\overline \Tr{^{\mu\nu}}=\frac{1}{2}\reg\,\Mg^{\,2-n}
\,\Big(\overline \Tr_{\!\mu\nu}\overline 
\Tr{^{\mu\nu}}-\frac{1}{n-2}\overline \Tr{^2}\Big)\,.\eqn{Lambdag}
\end{equation}
Finally the dilatonic contribution is obtained as
\begin{equation}
\widehat\Lambdar_{_{\rm D}}=\frac{1}{2}\widehat \phib\,
\overline \Tr =\frac{1}{2(n-2)}\reg\,
\Md^{\,2-n}\,\overline \Tr{^2}\,.\eqn{Lambdad}
\end{equation}
We should point out that in performing this calculation we have
included an extra factor $1/2$ which takes into account the double
counting implicit in the action renormalization procedure.

Using the defining relations (\ref{58},\ref{59}) and the conditions 
(\ref{72b}), it can be checked directly that the preceding prescriptions 
(\ref{Lambdaq},\ref{Lambdag},\ref{Lambdad}) actually do give rise to 
surface energy-momentum contributions of the forms given 
respectively by (\ref{38a},\ref{38c},\ref{38d}).
The renormalization of the action is thus
\begin{eqnarray}
\Delta\widehat\Lambdar&=&\frac{1}{4}\widehat{\hb}_{\mu\nu}\overline{\Tr}_{\mu\nu}
+\frac{1}{2}\widehat{\phib}\overline{\Tr}+\sum_q
\frac{1}{2q!}\widehat{\Br}_{\mu_1\dots\mu_q}\overline{\Jr}^{\mu_1\cdots\mu_q}
\nonumber\\
&=&\frac{\reg}{2(n-2)}\Bigg(\Mg^{2-n}\left[(n-2)\overline{\Tr}^{\mu\nu}\overline{\Tr}_{\mu\nu}
-\overline{\Tr}^2\right]+\Md^{2-n}\overline{\Tr}^2+\sum_q
\frac{\Mq^{2q+2-n}}{q!}\overline{\Jr}_{\mu_1\dots\mu_q}\overline{\Jr}^{\mu_1\cdots\mu_q}\Bigg)\,.
\end{eqnarray}
The validity of this coherence criterion means that the  ``dressed'' brane model
characterised by the action density, $\widetilde\Lambdar$,
will indeed have  a corresponding ``dressed'' surface stress momentum energy 
density tensor $\widetilde \Tr{^{\mu\nu}}$ of the required form, as given by 
(\ref{newT}).

\section{Maximal $(p+1)$-form and Nambu-Goto Brane}

An important special case, including that of an ordinary axionic
coupling to a string in 4-dimensional spacetime ~\cite{C99}, is that 
 for which the number of indices of the form field 
$\Br_{\mu_1...\mu_q}$ is equal to the dimension of the brane 
worldsheet. In this case, $q=p+1$, the corresponding source tensor on
the brane must simply be proportional to the corresponding surface
measure tensor, that is,
\begin{equation}
 \overline \Jr{^{\mu_0...\mu_p}}=\bar\kappar\, {\calE}^{\mu_0...\mu_p}
\end{equation}
with a proportionality coefficient $\bar\kappar$ having a uniform
value over the worldsheet in order for the conservation condition 
(\ref{fluxcons}) to be satisfied. It follows that the regularized 
action contribution will have the {\it constant} form
\be\widehat\Lambdar_{[p+1]}=-\frac{1}{2(n-2)}\reg\,\Mr_{[p+1]}^{\,2p+4-n}
\,\overline \kappar^2 \, .\fe
The manifestly negative definite nature of this axionic contribution is what
makes it possible in particular cases~\cite{C99} for it to be cancelled by 
the corresponding, manifestly positive definite, contribution
(\ref{Lambdad}) from the dilaton, as discussed below.

It is worthwhile to consider the
simplest dimensionally unrestricted application, which is to a NG
$p$-brane, that is, one for which the master
function $\Lambdar$ is just a constant, which we can express in terms
of a mass scale, as
\be \overline\Lambdar= -\Mk^{\,\rm p+1} \, ,\eqn{70}\fe
where $\Mk$ is a fixed mass scale that will be referred 
to as the Kibble mass to distinguish it from other mass scales in the theory.  
In the context of Superstring theory this quantity $\Mk$ 
is  usually supposed to be of the order of magnitude of the Planck mass 
$\Mp$, whereas in the context of cosmic string theory it
is generally expected that it should be of the same order of magnitude as the 
Higgs expectation value related to symmetry breaking.

In this special case, the energy-momentum tensor is of course simply 
proportional to the fundamental tensor,
\be \overline \Tr{^{\mu\nu}}=-\Mk^{\,p+1}\etag^{\mu\nu}  \, ,\eqn{NGem}\fe
so its trace will be given by
\begin{equation}
\overline \Tr=-(p+1)\Mk^{\,p+1}\,.
\end{equation}
The corresponding the hyper-Cauchy tensor is found~\cite{BC95} to be
\begin{equation}
\eqn{72} \overline{\calC}{^{\mu\nu\rho\sigma}}=\Mk^{\, \rm p+1}
\big(\etag^{\mu(\rho}\eta^{\sigma)\nu}-\frac{1}{2}
\etag^{\mu\nu}\eta^{\rho\sigma}\big)\quad\Longrightarrow\quad
\overline{\calC}{^{\mu\nu}}=\frac{1-p}{2}
\Mk^{\,p+1} \etag^{\mu\nu}\,,
\end{equation}
so it is apparent that $\overline{\calC}{^{\mu\nu}}$ will vanish in
the string case, $p=1$. The combination involved in the expression
(\ref{22b})
for the dilatonic force density will be given by
\be \overline \Tr{^{\mu\nu}}+\overline{\calC}{^{\mu\nu}}=
-\frac{p+1}{2}\Mk^{\,p+1}\etag^{\mu\nu}\, , \fe
which never vanishes.

The dynamical equation of motion (\ref{20})
can be seen from (\ref{17}) to reduce to the form
\be \Mk^{\,p+1} \Kg^\rho = - \fr{^\rho}\, ,\eqn{75}\fe
in which, by (\ref{22a}), the gravitational contribution to the surface force
on the right hand side can be written \cite{BC95} in the form
\begin{equation}
{\widehat\fr}_{_{\!\rm G}}{^\mu}=\Mk^{\,p+1}\left(
\perpg^{\mu\nu}\etag^{\rho\sigma}\Big( \nab_{\rho}\hb_{\nu\sigma}-
\frac{1}{2}\nab_{\nu}\hb_{\rho\sigma}\Big)+
\Big(\perpg^{\mu\nu}\Kg^{\rho}+\frac{1}{2}\etag^{\rho\nu}
\Kg^{\mu}-\Kg^{\nu\rho\mu}\Big)\hb_{\nu\rho}\right)
\,, \eqn{gravforce}
\end{equation}
while the corresponding dilatonic contribution will be given by
\be {\widehat\fr}_{_{\!\rm D}}{^\mu}=(p+1)\,\Mk^{\,p+1}\Big(
\phib\, K^\mu-\perpg^{\mu\nu}\nab_{\!\nu}\phib\Big)\,. \eqn{dilforce}\fe

Much of the early work on
perturbations of cosmic strings is flawed by the omission of both the
gradient terms and the orthogonal projection operator,
$\perpg^{\mu\nu}$, in (\ref{gravforce}).
One of the reasons the problem was not noticed in earlier studies of
NG string self-interactions is that, in the particular case when
$\hb_{\mu\nu}$ is due just to self interaction, the dominant
short-range contribution responsible for the divergence turns out to
be restricted in such a way as to give a result that does satisfy the
orthogonality requirement.
However, other forces, such as those due to external sources would
require the correct formulae (\ref{gravforce},\ref{dilforce}) to be used.

We can write out the full expression for the self-force acting on the brane as
\begin{equation}
\widehat\fr^\mu=\left[(p+1)(n-p-3)\Mk^{\,2p+2}\Mg^{\,2-n}
+(p+1)^2\Mk^{\,2p+2}\Md^{\,2-n}-\bar\kappar^2
\Mr_{[p+1]}^{\,2p+4-n}\right]\frac{\reg}{2(n-2)}\,\Kg^\mu\,,
\end{equation} 
and, the renormalization of the action as
\begin{equation}
\Delta\widehat\Lambdar=\left[(p+1)(n-p-3)\Mk^{2p+2}\Mg^{2-n}+
(p+1)^2\Mk^{2p+2}\Md^{2-n}-\bar\kappar^2\Mr_{[p+1]}^{\,2p+4-n}
\right]\frac{\reg}{2(n-2)}\,.
\end{equation}

\subsection{Previously derived special cases}

A number of special cases were discussed in section \ref{intro}. The
first was the point particle in four dimensions coupled to
electromagnetism, that is, $p=0$, $n=4$, $M_{\rm G}=M_{\rm D}=0$ in
which case 
\begin{equation}
\widehat f^{\mu}=-{{\bar\kappa}^2\over 4}\reg K^{\mu}\,,\qquad 
\Delta\widehat\Lambda=-{{\bar\kappa}^2\over 4}\reg\,,
\end{equation}
where ${\bar\kappa}=e$ the electromagnetic coupling,
$\reg=(\pi\epsilon)^{-1}$ and $\epsilon$ corresponds to the radius of
the electron. Another is the case of the global, or axion, string
which corresponds to $p=1$, $n=4$, $M_{\rm G}=M_{\rm D}=0$ and hence
\begin{equation}
\widehat f^{\mu}=-{{\bar\kappa}^2\over 4}M_{[2]}^2K^{\mu}\reg\,,\qquad 
\Delta\widehat\Lambda=-{{\bar\kappa}^2\over 4}M_{[2]}^2\reg\,,
\end{equation}
with ${\bar\kappa}=1$, $M_{[2]}=\sqrt{2\pi}f_{\rm a}$ and $\reg=4\log(\Delta/\epsilon)$.

The case of pure gravity ($M_{\rm D}=M_{[p+1]}=0$) was considered in
\cite{BCM04} and for this case the calculation presented here yields 
\begin{equation}
\widehat f^{\mu}=(p+1)(n-p-3)M_{\rm K}^{2p+2}M_{\rm G}^{2-n}\reg\,,\qquad 
\Delta\widehat\Lambda=(p+1)(n-p-3)M_{\rm K}^{2p+2}M_{\rm G}^{2-n}\reg\,,
\end{equation}
where, in the notation of \cite{BCM04}, $\lambda=M_{\rm K}^{p+1}$ and $G=M_{\rm G}^{2-n}$. 

\subsection{Co-dimension two}
\label{codim2}

The case of co-dimension two ($n=p+3$) is of interest since the
gravitational self-force is exactly zero. The dilatonic and
form-fields have opposite signs and their respective self-forces can
be made to cancel if 
\begin{equation}
{M_{\rm K}^2\over M_{\rm D}M_{[p+1]}}=\left({{\bar\kappa}\over p+1}\right)^{2\over p+1}\,.
\end{equation}
For the special case of $n=4$, $p=1$, this corresponds to the case previously discussed~\cite{DH,gibbons,CHH90,C99} where 
\begin{equation}
M_{\rm K}^2={1\over 2}{\bar\kappa M_{\rm D}M_{[2]}}\,.
\label{cancel}
\end{equation}
In the notation of \cite{CHH90} $\alpha=M_{\rm G}/M_{\rm D}$,
$\lambda={\bar\kappa}M_{\rm G}M_{[2]}/2$ and $\mu=M_{\rm K}^2$, and
hence the condition (\ref{cancel}) requires that $\alpha=1$ and
$\lambda=\mu$.

The co-dimension two case is also interesting when $n=6$ in the
context of brane-world models.  Many authors have observed that, for
a four-dimensional brane-world in six dimensions, the gravitational
effect of the bare tension of the brane is to produce a conical
deficit in the spacetime which is unobservable to a brane-based
observer.
Most of the work in the literature has considered specific solutions
and infinitely thin branes; our work has extended this to allow for
very general brane configurations which are extrinsically curved
within the spacetime~\cite{BCM04} and can account for branes of finite
thickness~\cite{CBU}.  This current paper includes the effect of the
dilaton and form-field effects which can cancel each other with an
appropriate choice of the mass scales $M_{\rm D}$, $M_{\rm K}$ and
$M_{[4]}$; this choice of coupling is the one selected naturally by
certain sting models, as described in~\cite{DH}.

The fact that the self-force is zero in the co-dimension two case has
obvious implications for the cosmological constant problem.  If a
vacuum energy component of matter on the brane does not gravitate, the
excessive values predicted by quantum field theories no longer
present a fine-tuning problem.  This is not a complete solution,
however, because the bulk could still induce a cosmological
acceleration on the brane.  Furthermore, cosmological inflation would
not be able to operate by the usual mechanism in such models.

\section{Discussion and conclusions}

We have calculated the self-force of branes due to classical
gravitational, dilatonic and form-field mediated interactions.
This analysis is very general, accounting for finite thickness effects
via the analysis in ~\cite{CBU} and allowing for the brane to be
curved provided that extrinsic curvature scale is long compared to the
ultra-violet cut-off associated with the brane thickness.  We have
also expressed these self-interactions as renormalizations of the
action.

For the gravitational force alone, the force is proportional to the
extrinsic curvature vector $K^\mu$.  In the special case of
codimension two, the force is actually zero, so the self-interactions
cancel.  This is a well-known result for cosmic strings in four
dimensions; the tension of the string determines the conical deficit
angle but does not affect the geometry away from the string or the
induced metric on the string.  More recently, this is has been
fashionable in the context of co-dimension two brane-worlds as a
possible resolution of the cosmological constant problem.

When a dilaton and a maximal $q$-form are included, certain
combinations of couplings result in zero total self-force.  Many
Superstring theories select precisely those couplings, suggesting that
our calculation could be used as a consistency relation for models,
i.e., that the self-force should always be zero for extended objects
in a consistent fundamental theory.

The formalism could be applied to many situations where there is an
extended object in a spacetime.  In particular it could be applied
to the Dirac--Born--Infeld (DBI) action which is of interest in string
theory when studying D-branes.

\acknowledgments

AM is supported by PPARC, and for much of this project was supported by Emmanuel College, Cambridge.

\end{document}